\documentstyle[prl,preprint,aps]{revtex}

\begin{document}
\title{Precise frequency measurements of the $D$-lines 
and fine-structure interval in K}
\author{Ayan Banerjee, Dipankar Das, and Vasant 
Natarajan\thanks{Electronic address: 
vasant@physics.iisc.ernet.in}}
\address{Department of Physics, Indian Institute of 
Science, 
Bangalore 560 012, INDIA}

\maketitle

\begin{abstract}
We use a diode laser locked to a Rb transition as the 
frequency reference along with a scanning Michelson 
interferometer to make precise measurements on the 
$D_1$ and $D_2$ lines of potassium. The Rb reference 
frequency is known with sub-MHz accuracy. We obtain the 
following values for the energy levels: 12 985.169 
60(13)(15) cm$^{-1}$ for the $4P_{1/2}$ state ($D_1$ 
line); 13 042.875 95(14)(15) cm$^{-1}$ for the 
$4P_{3/2}$ state ($D_2$ line); and 57.706 35(19)(15) 
cm$^{-1}$ for the $4P_{3/2}-4P_{1/2}$ fine-structure 
interval. The leading source of systematic error 
cancels in the determination of the interval. The 
errors represent about an order of magnitude 
improvement over tabulated values.
\end{abstract}

Semiconductor diode lasers bring several advantages to 
the field of laser spectroscopy, the principal ones 
being tunability and narrow spectral width 
\cite{WIH91}. By placing a single-mode diode in an 
external cavity and using optical feedback from an 
angle-tuned grating \cite{MSW92}, it can be made to 
operate at single frequency (single longitudinal mode) 
with linewidth of order 1 MHz and tunability over 
several nm. In addition, techniques such as 
saturated-absorption spectroscopy using counter-propagating pump 
and probe beams can help eliminate the first-order 
Doppler effect and allow narrow hyperfine transitions 
within a given atomic line to be resolved. It is thus 
possible to get an absolute frequency calibration of 
the laser by locking to such a transition. Recently, we 
have shown that a diode laser locked to an atomic 
transition in Rb can be used along with a scanning 
Michelson interferometer as a precision wavemeter to 
determine the unknown frequency of a tunable laser 
\cite{BRW01}.

In this Letter, we apply this technique to determine 
the absolute frequencies of the $D$ lines in potassium. 
The frequency of the reference laser is known with 
sub-MHz accuracy, and the unknown laser is tuned to the one 
of the $D$-line transitions in K with $\sim$MHz 
accuracy, therefore the frequencies are also determined 
with $\sim$MHz accuracy. More importantly, we can 
determine the fine-structure interval in the $4P$ state 
with this accuracy. Knowledge of fine-structure 
intervals is important for several reasons, {\it e.g.\ 
}in the study of atomic collisions, astrophysical 
processes, and relativistic calculations of atomic 
energy levels. Alkali atoms are particularly important 
because of their widespread use in ultra-cold collision 
studies, photoassociation spectroscopy, atomic tests of 
parity violation, and more recently in Bose-Einstein 
condensation \cite{KET99}. The species we have studied 
is K in which there has been renewed interest recently 
because laser cooling can be used to produce a quantum 
degenerate Fermi gas \cite{DEJ99}. But the technique is 
more general and applicable to other alkali atoms such 
as Li or Cs, alkali-like ions, and indeed any system 
where precise knowledge of energy levels is important. 
We achieve an accuracy of 0.00015 cm$^{-1}$, which is 
one to two orders of magnitude better than the typical 
accuracy of the energy-level tables \cite{MOO71}. 

The interferometer used for the ratio measurement has 
been described extensively in a previous publication 
\cite{BRW01}. For consistency with the terminology 
there, we will call the two lasers as ``reference'' and 
``unknown'', respectively. The basic idea is to obtain 
the ratio of the two laser wavelengths using a scanning 
Michelson interferometer where both lasers traverse 
essentially the same path. As the interferometer is 
scanned, the interference pattern on the detectors goes 
alternately through bright and dark fringes. Since both 
lasers traverse the same path, the ratio of the number 
of fringes counted after scanning through a certain 
distance is the ratio of the two wavelengths. The ratio 
obtained is a wavelength ratio in air, however, the 
wavelength ratio in vacuum (or equivalent frequency 
ratio) is easily calculated by making a small 
correction for the dispersion of air \cite{EDL66} 
between the reference wavelength and the unknown.

The reference laser is a diode laser system built 
around a commercial single-mode diode. The output is 
first collimated using an aspheric lens to give an 
elliptic beam of 5.8 mm $\times$ 1.8 mm $1/e^2$ 
diameter. The laser is then frequency stabilized in a 
standard external-cavity design (Littrow configuration) 
using optical feedback from an 1800 lines/mm 
diffraction grating mounted on a piezoelectric 
transducer \cite{MSW92}. Using a combination of 
temperature and current control, the laser is tuned 
close to the 780 nm $D_2$ line in atomic Rb. A part of 
the output beam is tapped for Doppler-free 
saturated-absorption spectroscopy in a Rb vapor cell. The 
injection current into the laser diode is modulated 
slightly to obtain an error signal and the laser is 
locked to the $5S_{1/2},F=2 \rightarrow 
5P_{3/2},F'=(2,3)$ crossover resonance. The frequency 
of this resonance has been measured previously to be 
384 227 981.877(6) MHz \cite{YSJ96}. The elliptic beam 
from the laser is directly fed into the Michelson 
interferometer. The large Rayleigh ranges ($\sim$34 m 
and $\sim$3 m in the two directions) ensure that the 
beam remains collimated over the length of the 
interferometer and diffraction effects are not 
significant.

The relevant energy levels of $^{39}$K are shown in 
Fig.\ 1. The ground state has two hyperfine levels with 
$F=1$ and 2. The $D_1$ line is the $4S_{1/2} 
\leftrightarrow 4P_{1/2}$ transition at 770.1 nm. The 
$4P_{1/2}$ state again has two hyperfine levels with 
$F=1$ and 2. The $D_2$ line is the $4S_{1/2} 
\leftrightarrow 4P_{3/2}$ transition at 767.7 nm. The 
upper state has 4 excited levels, with $F=0$, 1, 2, and 
3. The measurements on the $D$ lines of K are done 
using a single-frequency tunable Ti-sapphire laser 
(Coherent 899-21). The laser can be tuned from 700--800 
nm and is frequency stabilized to an ovenized reference 
cavity that gives it a linewidth of 500 kHz. A part of 
the output is split off for saturated-absorption 
spectroscopy in a K vapor cell. The cell is heated to 
about 70$^{\rm o}$C to get sufficient vapor. The 
saturated-absorption spectra on the $D_1$ and $D_2$ 
lines for transitions starting from the $F=2$ ground 
level are also shown in Fig.\ 1. The two hyperfine 
transitions in the $D_1$ line are well resolved, as 
seen in the figure. This is because the hyperfine 
interval in the $4P_{1/2}$ state is about 60 MHz. On 
the other hand, all the hyperfine levels in the 
$4P_{3/2}$ lie within a range of 30 MHz, and the 
individual transitions in the $D_2$ line are not 
resolved. However, the peak in the spectrum corresponds 
to the $F=2 \rightarrow F'=(2,3)$ crossover resonance, 
and this is what we use for the measurements. For 
transitions starting from the $F=1$ ground level, the 
peak corresponds to the $F=1 \rightarrow F'=(1,2)$ 
crossover resonance.

To get several independent measurements of the energy 
levels, we tune the Ti-sapphire laser to different 
hyperfine transitions of the $D_1$ and $D_2$ lines in 
K. In each case, the wavelength ratio is measured about 
600 times. The values are statistically analyzed to 
obtain the mean ratio and the $1\sigma$ (statistical) 
error in the mean. The mean wavelength ratios obtained 
from four such measurements (two for each line) are 
listed in Table I. We extract the hyperfine-free energy 
level of the state by making two corrections to the 
measured ratio. First, we convert the wavelength ratio 
in air to a frequency ratio using the refractive index 
of air from Edl\'en's formula \cite{EDL66}: $n=1.000$ 
275 163 at 780.0 nm (reference), $n=1.000$ 275 231 at 
770.1 nm ($D_1$), and $n=1.000$ 275 255 at 767.7 nm 
($D_2$). Then we remove the hyperfine frequency shifts 
shown in Fig.\ 1, which are known to sub-MHz accuracy 
\cite{AIV77}. The extracted values of the energy levels 
are also listed in Table I.

The main source of statistical error is that the 
frequency counter counts zero-crossings and does not 
count fractional fringes. The total number of fringes 
counted depends on the fringe rate (or cart speed) 
coupled with the 1 s integration time. We use about 20 
cm of cart travel per measurement, which results in a 
single-shot statistical error of about 5 parts in 
$10^7$ in each data set \cite{foot1}. The error of less 
than 2 parts in $10^8$ in Table I comes after averaging 
over $\sim 600$ individual measurements. We also plot a 
histogram of the individual points to make sure that 
the distribution around the mean is Gaussian. This 
guarantees that there is no significant statistical 
bias in the data.

There are several potential sources of systematic 
error, the main two being variation in the lock point 
of the lasers and non-parallelism of the two laser 
beams in the interferometer. We have checked for the 
first error by locking the unknown laser to different 
hyperfine transitions. The transitions are about 500 
MHz apart and the saturated-absorption spectrum in each 
case is quite different. Therefore, we expect the 
definition of the lock point also to be quite 
different. However, as seen from Table I, within the 
errors quoted the different ratios yield consistent 
values for the hyperfine-free energy level. This 
implies that there is no significant variation in the 
lock point of the lasers.

The second source of systematic error, namely that the 
two beams have a small angle between them, is more 
serious. Any misalignment would cause a systematic 
increase in the measured ratio given by $1/ \cos 
\theta$, where $\theta$ is the angle between the beams. 
We have tried to minimize this error in two ways. The 
first method is as follows. When the beams from the two 
arms of the interferometer are combined on the 
beamsplitter, two output beams are produced. Of these, 
the one on the opposite side of the beamsplitter from 
the input beam has near-perfect contrast and is 
detected for the measurement. The unused output beam 
(the one on the same side of the beamsplitter as the 
input beam) propagates towards the other laser, and can 
therefore be used as a tracer for checking the 
parallelism of the lasers. Using this method, we ensure 
that the reference and unknown beams are parallel over 
a distance of about 3 m \cite{foot2}. The second method 
to check for parallelism is to look for a minimum in 
the measured ratio as the angle of the unknown beam is 
varied. This works because the measured value is always 
larger than the correct value, whether $\theta$ is 
positive or negative, and becomes minimum when $\theta 
= 0$. We find this method to be more reliable and it 
quickly guarantees us that the beams are perfectly 
aligned.

It is difficult to get a quantitative estimate of the 
systematic error due to this non-parallelism. However, 
it is important to note that the error cancels in the 
determination of the fine-structure interval. This is 
because the measurements on the $D_1$ and $D_2$ lines 
are done without changing the alignment in the 
interferometer. The direction of the output of the 
Ti-sapphire laser does not change when its wavelength is 
changed, therefore any misalignment angle remains the 
same during all the measurements. If there is a 
misalignment, the measured frequencies on both the 
$D_1$ and the $D_2$ lines will be systematically higher 
but by the same amount, so that when we take their 
difference to obtain the fine-structure interval, the 
error will cancel.

As a further check on possible systematic errors in our 
measured frequencies, we have repeated the measurement 
scheme using a different reference laser. This set of 
measurements was done using a reference diode laser 
locked to a transition in the $D_1$ line of Rb at 795 
nm. The diode laser system is identical to the first 
one, except that after stabilization it is locked to 
the $5S_{1/2},F=3 \rightarrow 5P_{1/2},F'=3$ transition 
in $^{85}$Rb. The frequency of this transition is 377 
106 271.6(4) MHz \cite{BGR91}. Since this frequency is 
also known with sub-MHz accuracy, we do not expect any 
decrease in the precision of the measurements. However, 
because the laser is different, it requires completely 
new alignment of the interferometer, which gives us a 
good check on systematic errors arising from any 
misalignment angle between the beams. More importantly, 
it gives us a check on the reliability of converting 
wavelength ratios in air to vacuum using Edlen's 
formula because the reference wavelength has changed 
from 780 nm to 795 nm.

The results of the second set of measurements are 
listed in Table II. The values for the $D_1$ and $D_2$ 
lines of K are consistent with the earlier results in 
Table I, within the quoted errors. Therefore, we have 
strong reason to believe that our systematic errors are 
less than the statistical errors. Combining the results 
from Tables I and II, we obtain the following average 
values
\begin{eqnarray*}
4P_{1/2}:  12 \;\, 985.169 \;\, 60(13)(15) {\rm \ \ 
cm^{-1},} \\ 
4P_{3/2}:  13 \;\, 042.875 \;\, 95(14)(15) {\rm \ \ 
cm^{-1},} 
\end{eqnarray*}
and a precise value for the fine-structure interval
\[ 4P_{3/2}-4P_{1/2}:  57.706 \;\, 35(19)(15) {\rm \ \ 
cm^{-1}.} \]
The errors are statistical and systematic, 
respectively.

The values we obtain can be compared to the values 
listed in the K energy-level tables released by the 
National Institute of Standards and Technology 
\cite{MOO71}: 12 985.170 cm$^{-1}$ for the $4P_{1/2}$ 
state; 13 042.876 cm$^{-1}$ for the $4P_{3/2}$ state, 
and 57.706 cm$^{-1}$ for the fine-structure interval. 
Our results are consistent with these values but have 
significantly higher accuracy, thus demonstrating the 
power of our technique to improve the accuracy of the 
existing energy-level tables.

In conclusion, we have demonstrated that a diode laser 
stabilized on a Rb transition can be used as an 
absolute frequency reference to make precise 
measurements of the energy levels in atoms. The 
reference laser and a laser tuned to the atomic 
transition of interest are fed into a scanning 
Michelson interferometer to obtain their frequency 
ratio very precisely. We have used this technique to 
make frequency measurements on the $D_1$ and $D_2$ 
lines of K, and obtain improvement of about an order of 
magnitude over existing values. The leading source of 
systematic error is non-parallelism between the two 
laser beams in the interferometer. While we have 
several checks to minimize this error, it cancels to 
first order in the determination of the fine-structure 
interval. The technique should prove useful in other 
atomic systems and particularly alkali atoms, where 
transitions are easily accessible with tunable diode 
lasers and knowledge of fine-structure intervals is 
important. Since atomic energy levels are generally 
known to about 0.01 cm$^{-1}$ accuracy, our precision 
wavemeter \cite{BRW01} has the potential to improve 
these values by up to two orders of magnitude.

This work was supported by research grants from the 
Board of Research in Nuclear Sciences (DAE), and the 
Department of Science and Technology, Government of 
India.

\begin{figure}
\caption{
Energy levels of $^{39}$K. The figure shows the 
relevant energy levels of $^{39}$K in the ground $4S$ 
state and first excited $4P$ state. The various 
hyperfine levels are labeled with the value of the 
total angular momentum $F$, and the number on each 
level is the energy displacement (in MHz) from the 
unperturbed state. The two insets on the left-hand side 
are saturated absorption spectra on the $D_1$ and $D_2$ 
lines for transitions starting from the $F=2$ ground 
level.
}
\end{figure}

\begin{table}
\caption{ 
The table lists the measured wavelength ratios and the 
energy levels in K. The reference laser was locked to 
the $5S_{1/2},F=2 \rightarrow 5P_{3/2},F'=(2,3)$ 
crossover resonance in the $D_2$ line of $^{87}$Rb, 
corresponding to a frequency of 384 227 981.877(6) MHz. 
The unknown laser was tuned to various hyperfine 
transitions of the $D_1$ and $D_2$ lines in $^{39}$K, 
as listed. The hyperfine-free line wavenumber in vacuum 
was extracted by first correcting for the dispersion of 
air and then removing the hyperfine shifts shown in 
Fig.\ 1. The errors are statistical $1 \sigma$ 
deviations.
}
\begin{tabular}{ccc}
Measured transition & Wavelength ratio & Energy (cm$^{-1}$) \\
\tableline
\vspace*{-2mm}\\
$D_1$: $F=1 \rightarrow F'=(1,2)$ & 1.013 163 848(18) & 
12 985.169 64(23) \\
$D_1$: $F=2 \rightarrow F'=(1,2)$ & 1.013 162 640(19) & 
12 985.169 56(24) \\
$D_2$: $F=1 \rightarrow F'=(1,2)$ & 1.017 666 373(19) & 
13 042.875 93(24) \\
$D_2$: $F=2 \rightarrow F'=(2,3)$ & 1.017 665 211(16) & 
13 042.875 93(20) \\
\end{tabular}
\end{table}

\begin{table}
\caption{ 
Similar to Table I but now the reference laser was 
locked to the $5S_{1/2},F=3 \rightarrow 5P_{1/2},F'=3$ 
hyperfine transition in the $D_1$ line of $^{85}$Rb, 
corresponding to a frequency of 377 106 271.6(4) MHz. 
The unknown laser was tuned to hyperfine transitions of 
the $D_1$ and $D_2$ lines in $^{39}$K, as before.
}
\begin{tabular}{ccc}
Measured transition & Wavelength ratio & Energy (cm$^{-1}$) \\
\tableline
\vspace*{-2mm}\\
$D_1$: $F=2 \rightarrow F'=(1,2)$ & 1.032 296 511(16) & 
12 985.169 59(20) \\
$D_2$: $F=2 \rightarrow F'=(2,3)$ & 1.036 884 083(24) & 
13 042.876 03(30) \\
\end{tabular}
\end{table}


\begin{references}
\bibitem{foot1}
The calculated error from missing one fringe over a 
distance of 20 cm is about 1 part in $10^6$ at 780 nm. 
Our error is smaller because we are measuring a 
wavelength ratio close to 1 and chances are high that 
either both counters see a zero crossing or miss it.
\bibitem{foot2}
Ideally, the unknown beam should enter exactly where 
the reference beam exits so that both beams traverse 
exactly the same path, albeit in opposite directions. 
However, this causes one laser to feed into the other 
and destabilize it. Therefore we align the beams with a 
small offset and check for parallelism over a finite 
distance.
\end{references}
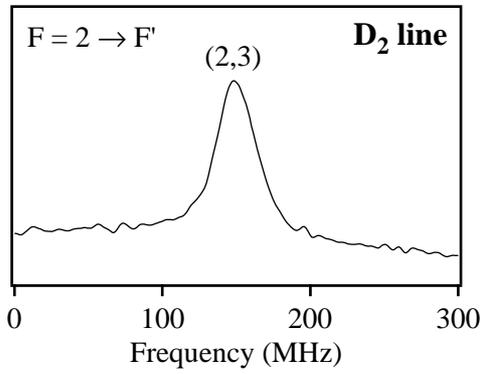
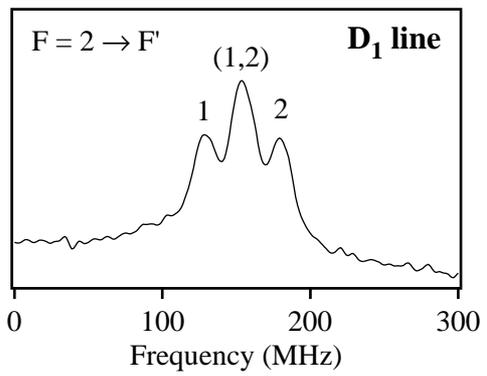
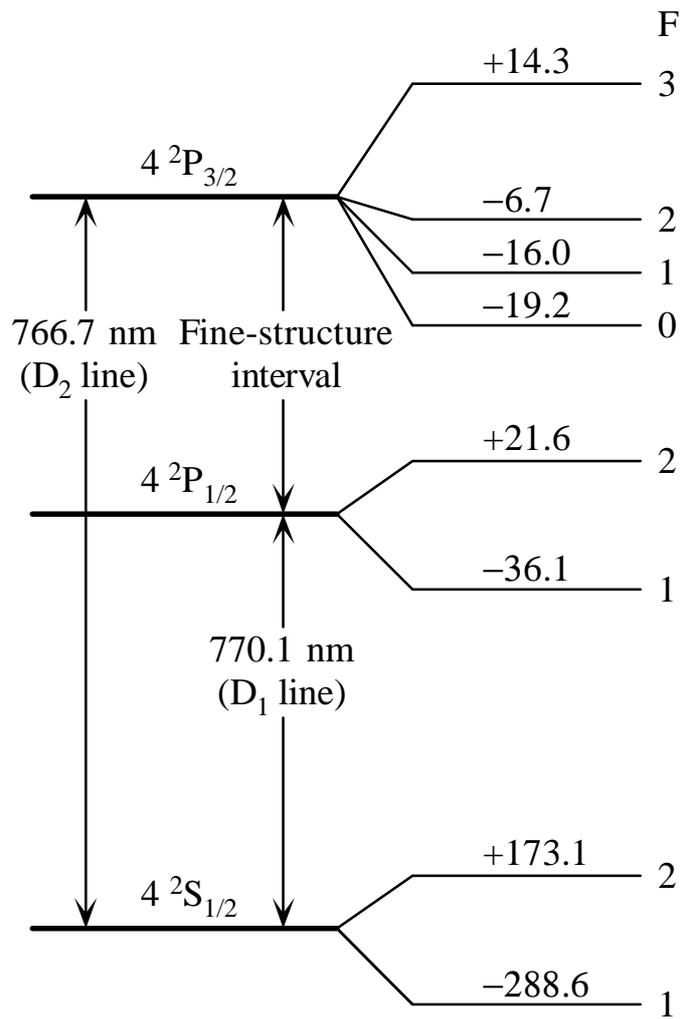
\end{document}